# Modeling the registration efficiency of thermal neutrons by gadolinium foils

D.A.Abdushukurov, M.A.Abduvokhidov, D.V.Bondarenko, Kh.Kh.Muminov, T.A.Toshov, D.Yu.Chistyakov

Physical-Technical Institute of the Academy of Sciences of the Republic of Tajikistan,

299/1, Ayni Ave, Dushanbe, Republic of Tajikistan,

E-mail: abdush@tajik.net

Abstract: In the paper we present the results of mathematical modeling of the registration efficiency of thermal neutrons for the converters made of natural Gd and its 157 isotope plane-parallel foils. In the performed calculations four fixed energies of neutrons with the corresponding wavelengths of 1, 1.8, 3 and 4 $A^0$ are taken into account. We calculate the efficiencies of the converter for electron escapes to frontward and backward hemispheres and their sum, depending on thickness of converting foils. Results of comparison of our calculations with the experimental data are presented.

Contents

**1. Introduction**

Converters of neutron radiation play the crucial role during development of detectors because they determine the basic characteristics of detectors, i.e. – the efficiency of registration, as well as both energetic and spatial resolution of the detector.

Gase-filled and solid-state converters of slow neutrons are widely used now. Several types of nuclear reactions are used in these converters. The most well-known among them are $^3$He and $^{10}$BF$_3$ converters. It should be mentioned that gas-filled converters are used in the wide class of gas-filled detectors [1],[2],[3]. Solid-state converters are used in gas-filled as well as in both scintillation and semiconductor detectors. But solid-state converters are rarely used in the detectors of neutrons; this is related first of all to their low efficiency of registration. Thus, the registration efficiency of detectors based on $^{10}$B and $^6$Li as usual does not exceed 3-4 % and 1 %, respectively (see [1], [4]).

Among the solid-state converters of thermal neutrons, the highest registration efficiency up to 30 % is received using converters on the basis of gadolinium and especially its 157 isotope [5], [6]. In the paper [7] simulation of the registration efficiency of thermal neutrons by the gadolinium foil converters was performed. The following conditions were considered in this simulation: neutron flux (with 2 fixed energies) perpendicularly fell on to the surface of the converting foil of various thicknesses; the secondary electron escapes to the frontward and backward hemispheres were calculated. As a result, the detector on the basis of multi-wire proportional chamber with gadolinium converter has been developed. However, this kind of detector does not possess well enough spatial resolution; it is just about 10 mm. In order to improve the spatial resolution the same authors suggested to use the converter-collimator, which was a sandwich of alternating lead foils interlaid by glass textolite with holes drilled in them by fine steps. This converter was pressed against a gadolinium foil and thus it limited the electron tracks in gas, simultaneously serving as an emitter of secondary electrons. It helped to improve the spatial resolution up to 2 mm, but the spatial resolution has turned out to be modulated by the step of apertures, moreover the efficiency of registration has sharply fallen [8]. Further, during the development of multi-step avalanche chambers with the gadolinium converters the spatial resolution of detectors was improved significantly (up to 1 mm) without using Jeavons converter [9].

Recently many authors were engaged in the development of detectors on the basis of gadolinium converters [10],[11],[12],[13],[14],[15],[16]. Simulations of the registration efficiency of thermal neutrons are also performed [17],[18],[19]. This fact testifies a significant interest to the gadolinium converters.

It is worth to mention that modeling of the processes of neutron capture by gadolinium nuclei recently found it application also in beam therapy in order to decrease dozes of radiation received by oncological patients during their treatment. Recently, since 2000, the new direction in oncology and treatment of cancer tumors has started to develop, which is called gadolinium neutron capture therapy (GdNCT) [20]. The method is based on the injection of gadolinium nuclei to the organism of patient in the composition of medicinal products and their selective absorption by the malignant tumor cells. Thus gadolinium nuclei, which have an extraordinary high section of interaction with thermal neutrons, allow to localize influence of radiation in the area of cancer cells. Therefore the basic radiation exposure is provided by both low-energy internal conversion electrons and Auger electrons.

In our previous calculations of the registration efficiency of Gd-based detectors [17] we used data on internal conversion coefficients from the paper [21]. There are more precise new data now [22]. Analysis of these data had shown that we could improve our calculations and obtain more accurate results. This fact stimulated us to perform new calculations.

In this work we will carry out model calculation of the efficiency of registration of thermal neutrons by the foil converters made of natural gadolinium and its 157 isotope. Also we compare our obtained modeling data with the experimental results.

## 2. Physical basis and model approximation

In the process of the capture of thermal neutrons by gadolinium nuclei, in addition to the radioactive γ-quanta, here both internal conversion electrons and Auger electrons are being emitted. These are the electrons that are mainly registered by the position sensitive detectors and especially by the gas-filled ones, which have low γ-quanta registration efficiency. Therefore, in our calculations, we take into account only those electrons, which arise as a result of neutron conversion on gadolinium nuclei.

As it is well-known, natural gadolinium is a mixture of isotopes that could participate in the (n, γ) nuclear reaction. Table 1 summarizes the main characters of Gd isotopes, including their cross sections of interactions with neutrons, daughter isotopes and half-decay periods of unstable daughter isotopes [23].

Table 1

| Isotope | Abundance | Cross section (b) | Daughter isotope | $T_{1/2}$ |
|---|---|---|---|---|
| $^{nat}$Gd | 100 | 48890 | - | - |
| $^{152}$Gd | 0.2 | 1100 | $^{153}$Gd | 241.6 d |
| $^{154}$Gd | 2.2 | 90 | $^{155}$Gd | Stable |
| $^{155}$Gd | 14.7 | 61000 | $^{156}$Gd | Stable |
| $^{156}$Gd | 20.6 | 2.0 | $^{157}$Gd | Stable |
| $^{157}$Gd | 15.68 | 255000 | $^{158}$Gd | Stable |
| $^{158}$Gd | 24.9 | 2.4 | $^{159}$Gd | 18.6 h |
| $^{160}$Gd | 21.9 | 0.8 | $^{161}$Gd | 3.66 min |

As one can see in this table, the most interesting for our calculations are natural Gd and its 155 and 157 isotopes, which have abnormally high cross sections of interaction with neutrons. Other isotopes give an insignificant contribution to the interaction with neutrons.

In this paper we will consider natural gadolinium and its 157 isotope only.

In the reaction of $^{157}$Gd neutron capture, 7937.33 keV energy is emitted. In total 390 lines with energy ranges from 79.5 up to 7857.670 keV with line intensity of $2*10^{-3}$ up to 139 gamma-quanta on 100 captured neutrons are emitted. In the Table 2 the most intensive, low - energy gamma-lines having high coefficient of internal conversion are presented. In the fig. 1 the histogram showing dependence of gamma quantum intensity on the energy is presented [24].

Table 2

| Isotope | Daughter isotope | Eγ [kev] | Cross section [b] | Iγ [1/100 n] |
|---------|------------------|----------|-------------------|--------------|
| 157-Gd | 158-Gd | 79.510 | 4010(100) | 77.3(19) |
| 157-Gd | 158-Gd | 135.26 | 38(4) | 0.73(8) |
| 157-Gd | 158-Gd | 181.931 | 7200(300) | 139(6) |
| 157-Gd | 158-Gd | 212.97 | 10.8(7) | 0.21(13) |
| 157-Gd | 158-Gd | 218.225 | 55(4) | 1.06(8) |
| 157-Gd | 158-Gd | 230.23 | 20.0(11) | 0.385(21) |
| 157-Gd | 158-Gd | 255.654 | 350(19) | 6.7(4) |
| 157-Gd | 158-Gd | 277.544 | 493(12) | 9.50(23) |
| 157-Gd | 158-Gd | 365 | 59(5) | 1.14(10) |
| 157-Gd | 158-Gd | 780.14 | 1010(22) | 19.5(4) |
| 157-Gd | 158-Gd | 944.09 | 3090(70) | 59.5(13) |
| 157-Gd | 158-Gd | 960 | 2050(130) | 39.5(25) |
| 157-Gd | 158-Gd | 975 | 1440(21) | 27.8(4) |

Since, there are low-energy gamma quanta in the spectrum, thus during their emission the probability of an electron irradiation from the atom shell (electrons of internal conversion) is very high. Nucleus eliminates its excitation by the irradiation of gamma-quantum, but it also can emit the close located electrons. As usual the K-electron (electron from the K-shell) is emitted, but an electron from the higher shell (like L, M, N and so on) also could be emitted. Created electron vacancy (the electron hole) is filled by another electron from the higher level. This process is accompanied by X-ray irradiation, or by the irradiation of Auger electron.

The effect of internal conversion is accompanied by the significant X-ray radiation, which could positively affect on the use of scintillation detectors with the fine-dispersed gadolinium. We will only consider electrons with the energies higher than 20 keV.

Data on coefficients of internal conversion are different in various sources [21] that lead to the divergences in quantity of the secondary electrons. In our last modeling, we based on the last data presented in the database [22].

In the Table 3 the most intensive lines of electrons are given, whose probability of emission exceeds 0.03/100 neutrons, for the energies of primary gamma quantum less than 1 MeV. Data on Auger electrons which are formed during the K-shell filling are presented as well. The maximum number of formed vacancies on K-shell (electron holes) does not exceed 45.2 to 100 falling neutrons.

Table 3

| Electron Energies (keV) | Electron output 1/100 n | Electron path in Gadolinium (μm) | Energy of primary gamma quantum | Comment, level |
|---|---|---|---|---|
| 29.3 | 35,58 | 4,7 | 79.51 | K |
| 34.9 | 7.9 | 6,29 | | K- Auger |
| 71.7 | 5,57 | 20,7 | 79.51 | L |
| 78 | 1,2 | 23,78 | 79.51 | M |
| 131.7 | 6,96 | 55,70 | 181.93 | K |
| 174.1 | 0,99 | 86,27 | 181.93 | L |
| 180.4 | 0,21 | 91,23 | 181.93 | M |
| 205.4 | 0,14 | 111,47 | 255.66 | K |
| 227.3 | 0,16 | 130,27 | 277.54 | K |
| 729.9 | 0,03 | 649,38 | 780.14 | K |
| 893.85 | 0,06 | 830,05 | 944.09 | K |
| 911.8 | 0,04 | 849,83 | 960 | K |
| 926.8 | 0,03 | 866,35 | 975.4 | K |

In our calculations, totally 444 discrete electron energies with the output probability of more than $10^{-5}$ on 100 falling neutrons were considered. On fig. 2 the histogram is presented, which shows the dependence of the most intensive lines of electron intensity on their energy.

The most important characteristic of converter is the probability of secondary electron output, emitted during the irradiative capture of neutrons by the substance of converter. Since, generally, electrons have low energies and small paths in the substance of converter it puts forward the additional requirements to the thickness of the converter.

The absorption coefficient $F_0$ characterizes probability of absorption of electrons in the substance. If X is thickness of the converter, Re is an electron path in the substance of converter, then

$$F_0(X) = 1 - X * \rho / Re \qquad (1)$$

where $\rho$ is density of the substance of converter. For gadolinium it is $\rho = 7{,}9$ g/cm$^3$.

Fig. 3 shows the dependence of the electron path in gadolinium (g/cm$^2$) on their energy [25]. Size Re is defined both by the electron energy and by the specific ionization loss value

$$Re = \int_0^E dE(-dE/dX) \ . \qquad (2)$$

Attenuation of narrow collimated neutron beam in a thin layer of the substance is governed by the exponential law

$$F_x = F_0 \exp(-N_A \sigma X) \qquad (3)$$

where $F_x$ and $F_0$ are the neutron flux density after and before its passage through the layer of the material with the thickness X, respectively, $N_A$ is the number of nucleus in the volume of 1 cm$^3$, $\sigma$ is full microscopic cross-section of neutron interaction with the nuclei of substance.

In our calculations we will use four fixed neutron energies corresponding to thermal area. These are neutrons with wavelengths of 1, 1.8, 3 and 4 A$^0$. In table 4 their wavelengths, corresponding energies (eV) and velocities (m/s) are shown.

Table 4

| Wave Length A$^O$ | Energy of Neutrons (eV) | Velocity of Neutrons (m/s) | Capture Cross Section (barns) for NatGd | Capture Cross Section (barns) for 157Gd |
|---|---|---|---|---|
| 1 | 0,081894 | 3955 | 13 563.56 | 75 323.47 |
| 1,8 | 0,025276 | 2197,2 | 48 149.41 | 253 778.40 |
| 3 | 0,0090993 | 1318,3 | 70 597.77 | 367 842.60 |
| 4 | 0,0051184 | 988,76 | 89 066.84 | 464 373.40 |

Fig. 4 depicts the dependence of cross-section (barn) on the energy of incident neutrons (eV), for natural gadolinium and the same dependence for its 157 isotope [26]. Arrows indicate energy of neutrons for which we will carry out our calculations. As one can see, with the reduction of energy of neutrons the cross-section of interaction strongly increases. Especially it increases in the region of cold and ultra cold neutrons.

Modeling was realized by examining the simplest targets, namely the plane-parallel foils. The calculations for thermal neutrons with the fixed energies, which correspond to neutron wavelength of 1, 1.8, 3 and 4 $A^0$ were carried out; also both the thickness of converters from 1 µm to 40 µm, and isotopic composition of converter (for natural Gd and $^{157}$Gd) were varied.

In the calculation all electrons (appeared as a result of neutron capture act) are taken into account which are able to escape an infinite plane-parallel plate of the converter. The ratio of the number of electrons escaped from the foil to that of the incident ones is referred to as the efficiency of the converter.

Conventionally we divide a foil into more thin layers. For each elementary layer we count the probability of absorption of neutrons with the fixed energy. Efficiency of the converter will be determined by the sum of probabilities of neutron absorption and probability of the electron escape from the converter. In order to calculate the electron escape we have chosen a simple geometrical model. The choice of the model is made from the following assumptions: all electron emissions are isotropic, the length of path for any fixed energy of electrons ($Re_i$) is constant (fluctuation of power losses in the end of path is neglected). Then the density of probability to find electrons in the substance forms a sphere with the radius equal to $Re_i$ (for any fixed energy). If the center of the sphere is crossed by the plane, two identical hemispheres are formed, which correspond to the electron escape to the forward and backward hemisphere, thus the area of hemispheres could be considered as the probability of an electron output. In this case total probability is 100 %, and escape to the one of hemispheres is 50 %. If we begin to cross a sphere with a step much less than $Re_i$, segments will be formed whose area will be equal to the probability of electron escape. The step of iterations should be at least 100 times less than $Re_i$, and then the electron absorption under the big corners could be neglected. The area of a segment and therefore the probability of electron escape becomes equal to zero in the intersection of a sphere by the plane at the distance $Re_i$. The sum of probabilities of electron escapes for all energies, taking into account their weight contributions will determine the total electron escape probability.

In our calculations, the probability of neutron absorption for each elementary layer, as a result of reaction of inelastic interaction (n, $\gamma$), is determined, using database for fixed neutron energies. Both the probability of electron emission and probability of their escape from the body of the converter were calculated as well.

Results of calculations of neutron absorption in converters made from natural gadolinium and its 157 isotope are shown in Fig. 5. Figure shows that at 30 microns thickness of the natural gadolinium converter there takes place a full absorption of neutrons with the wave length $\succ$ 1.8 $A^O$. For its 157 isotope for neutrons with the wave lengths $\succ$ 1.8 $A^O$ the same absorption happens at 8 microns thickness of the converter.

In order to detect cold neutrons by the 157Gd converters we could limit ourselves by the thickness of 10 microns of the converter, if there no technological restrictions. It should be taken into account that ranges of the majority of electrons emitted in the reaction of radiating capture of neutrons do not exceed 5 microns. If we use natural gadolinium converter the situation is little bit more complex, since low absorbing ability results in the necessity to use of more thick (up to 20-40 micron) converters.

Obtained data on effectiveness and optimal thicknesses of converters for natural gadolinium and its 157 isotopes (see table 5 and 6 respectively).

Table 5

| $\lambda$ ($A^0$) | Efficiency E and optimal thickness T ($\mu$m) | | |
| --- | --- | --- | --- |
| | Forward | backward | total |
| 1 | 0.080 (16) | 0.107 (30) | 0.180 (22) |
| 1.8 | 0.145 (9) | 0.205 (19) | 0.328 (12) |
| 3 | 0.169 (6) | 0.232 (13) | 0.388 (9) |
| 4 | 0.190 (4) | 0.254 (12) | 0.415 (8) |

Table 6

| $\lambda$ (A⁰) | Efficiency E and optimal thickness T ($\mu$m) | | |
| --- | --- | --- | --- |
| | Forward | backward | total |
| 1 | 0.170 (5) | 0.244 (14) | 0.400 (8) |
| 1.8 | 0.265 (3) | 0.319 (6) | 0.569 (4) |
| 3 | 0.293 (3) | 0.336 (5) | 0.611 (3) |
| 4 | 0.305 (2) | 0.341 (4) | 0.633 (2) |

Fig. 6 and 7 shows the dependence of registration efficiency on converter thickness for neutrons with the different energies.

Our obtained results of calculations are compared with the experimental data presented in paper [18]. In this paper experimental data on the detection efficiency was measured in backward direction for six different energies and compared to a calibrated ³He counter. Enriched up to 90.5% ¹⁵⁷Gd converter was used. Works were performed in the reactors of Atominstitut in Vienna (ATI) and the ILL Grenoble.

The results of comparison are shown in the Fig. 8. For natural gadolinium our results coincide with the experimental data well enough, in the limits of errors. Errors of calculations are caused both by the precision of determining of gamma-quanta output (see Table 2) and by determining of neutron cross-section. This good consent testifies the correct choice of our models and performed calculations.

For the converter made of 157 isotope of Gd the experimental curve lays little bit higher. Our data are in good agreement with the theoretical limit, determined by the coefficients of internal conversion, which makes no more than 65 % for an isotropic electron escapes, i.e. no more than 32.5 % for the escape to one of hemispheres. If we take into account Auger electrons, which could also escape in other directions with respect to internal conversion electrons, the efficiency of the converter, could be improved by 4%, i.e. theoretical limit is 36.5 % for the electron escapes to the backward direction.

3. Conclusion.

Model calculations of the efficiency of registration of thermal neutrons by foil converters made of natural gadolinium and its 157 isotope are carried out. Processes of neutron absorption in the substance of converter and the probability of secondary electron escapes were examined. Calculations are performed for converters with the various thicknesses. We have chosen the most optimal converter thicknesses, both from natural gadolinium and its 157 isotope. While using converters from natural gadolinium it is possible to obtain total effectiveness of 18, 32, 38, 42%, respectively for the neutrons with the wavelengths of 1, 1.8, 3 and 4 $A^0$ for the converter thickness of 22, 12, 9, 8 microns. For the converters made of 157 isotope it is possible to reach total effectiveness of registration up to 40, 57, 61, 63% for the neutrons with the wavelengths 1, 1.8, 3 and 4 $A^0$ with the thickness of converter 8, 4, 3, 2 microns. Obtained results of calculation are compared with the experimental data. Our data for natural gadolinium in the limits of errors coincide with the experimental [18] data, which testifies the correctness of chosen models and algorithms of calculations. For the converters made of $^{157}$Gd our data are little bit lower, but nevertheless they do not exceed the maximal possible values.

Description of the figures

Fig. 1. Histogram of dependence of intensity of gamma-quanta on energy
for the reaction of $^{157}$Gd (n, gamma) $^{158}$Gd (by 100 neutrons).

Fig. 2. Intensity of electrons emitted in the reaction $^{157}$Gd (n, $\gamma$) $^{158}$Gd
depending on their energy (by 100 neutrons).

Fig. 3. Dependence of the paths of electrons on their energies
in gadolinium (g/cm$^2$).

Fig. 4. Cross-section of thermal neutrons capture, for the reaction (n, $\gamma$),
depending on energy of neutrons for natural gadolinium and its 157 isotope.

Fig. 5. A curve of neutron absorption with wave lengths 1, 1.8, 3, 4 A$^O$
in natural gadolinium and it s157 isotope.

Fig. 6. The dependence of the neutron registration efficiency of the natural gadolinium
converter on its thickness. Curve 1 and 2 correspond to emission of electrons into the
front and back hemisphere, respectively, curve 3 is their sum.

Fig. 7. The dependence of the neutron registration efficiency of the $^{157}$Gd converter
on its thickness. Curve 1 and 2 correspond to emission of electrons into the front
and back hemisphere, respectively, curve 3 is their sum.

Fig. 8. Comparison of our calculated data with the experimental data for backward escape
geometry. We use the experimental data obtained in Atominstitut in Vienna (ATI) and the
ILL Grenoble [18].

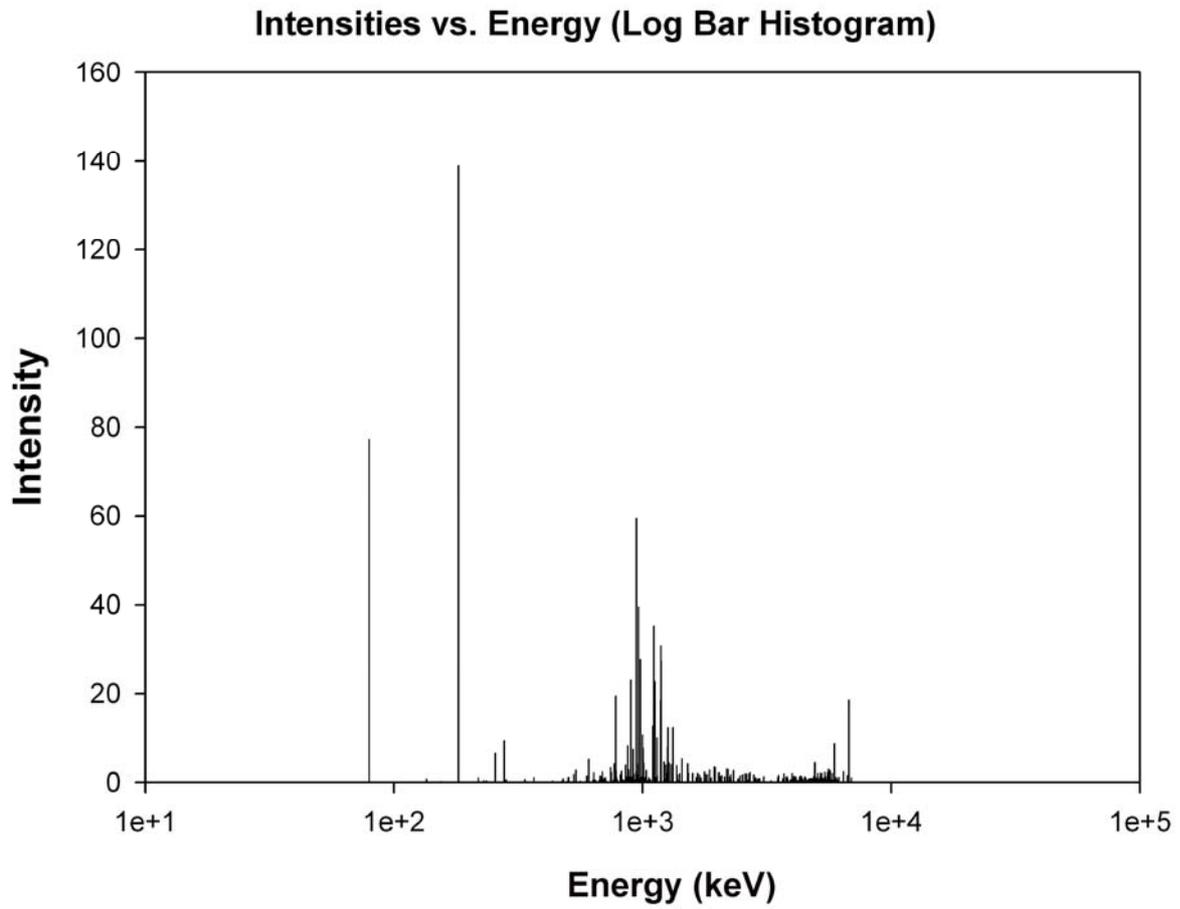

Fig. 1

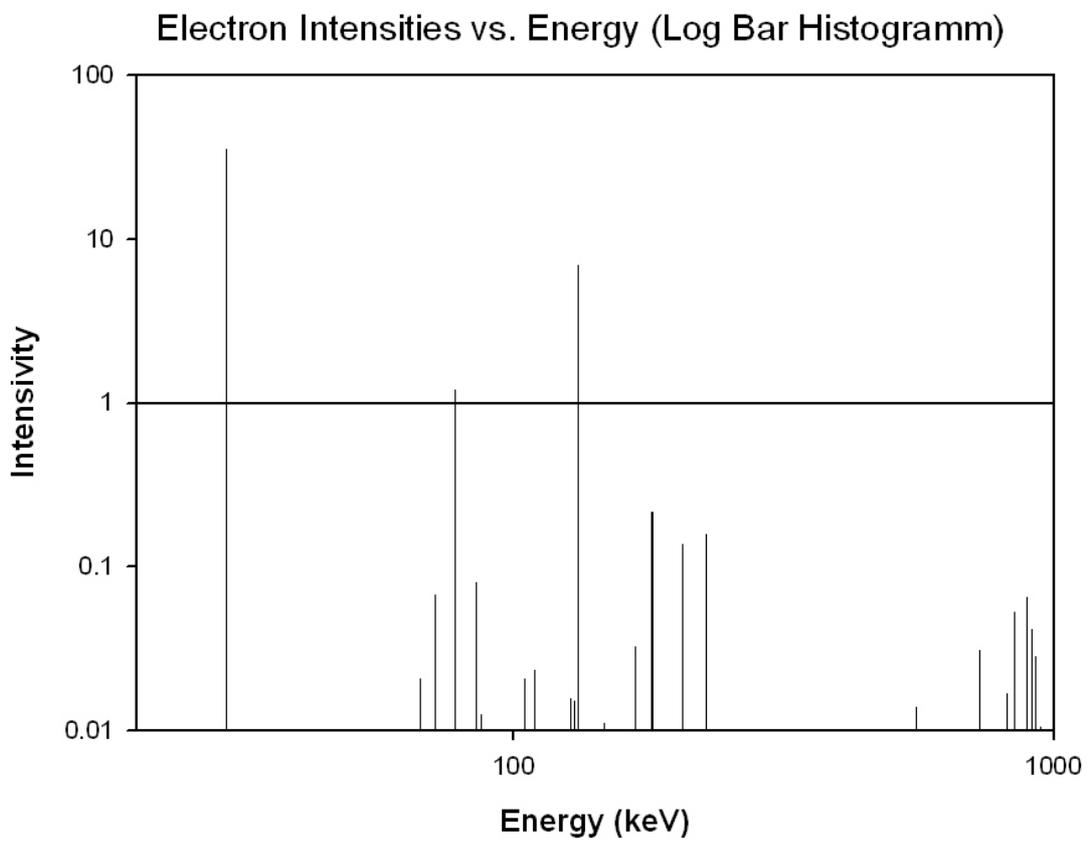

Fig. 2

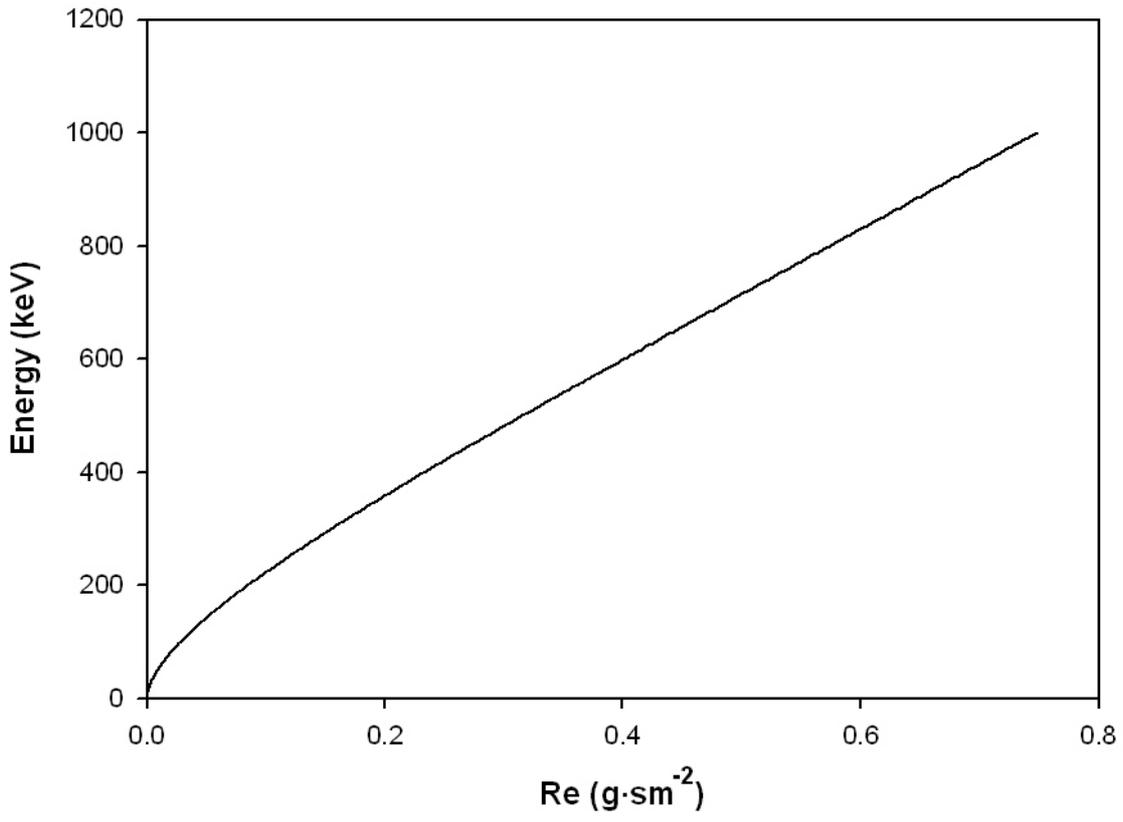

Fig. 3

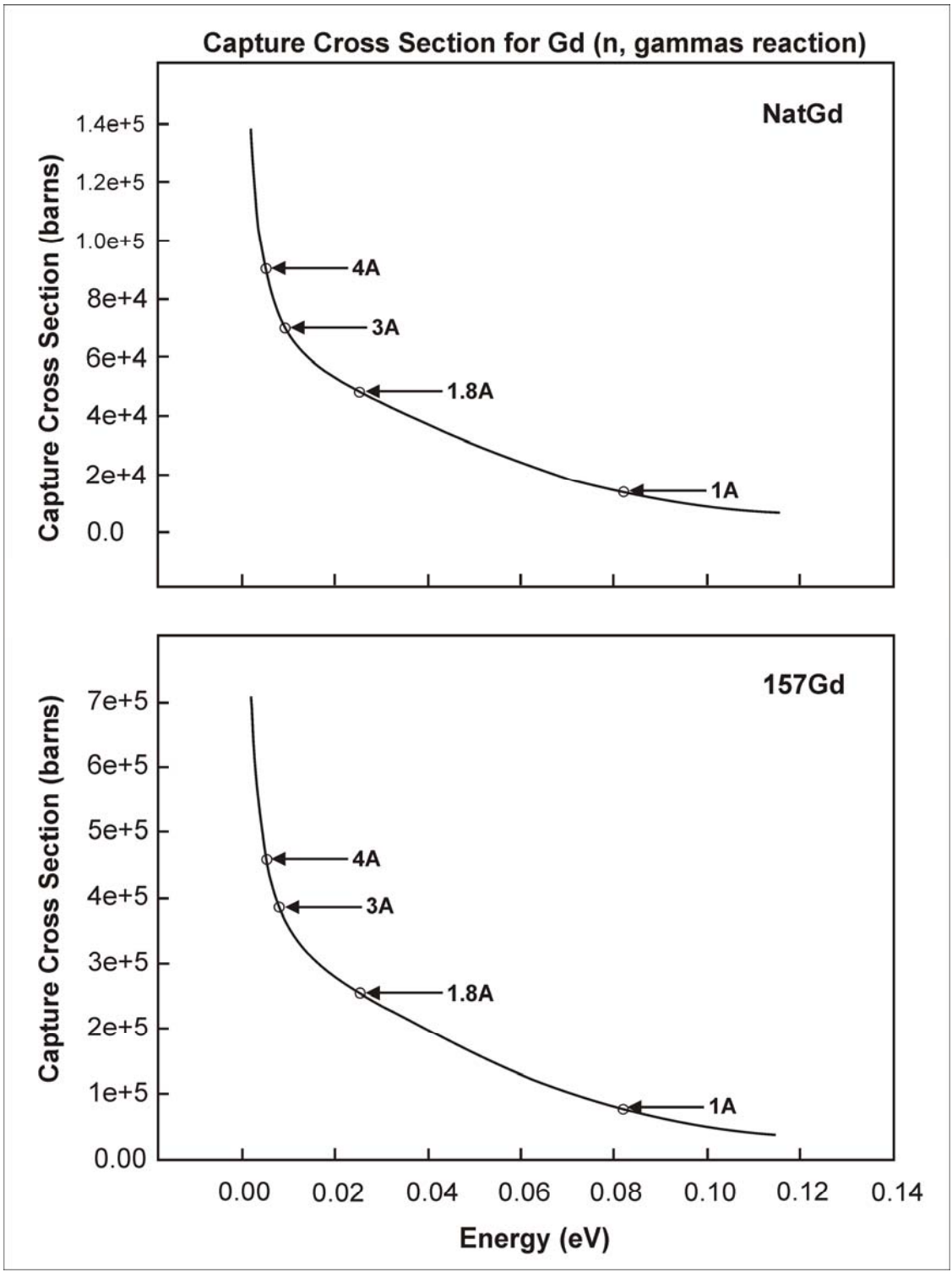

Fig. 4

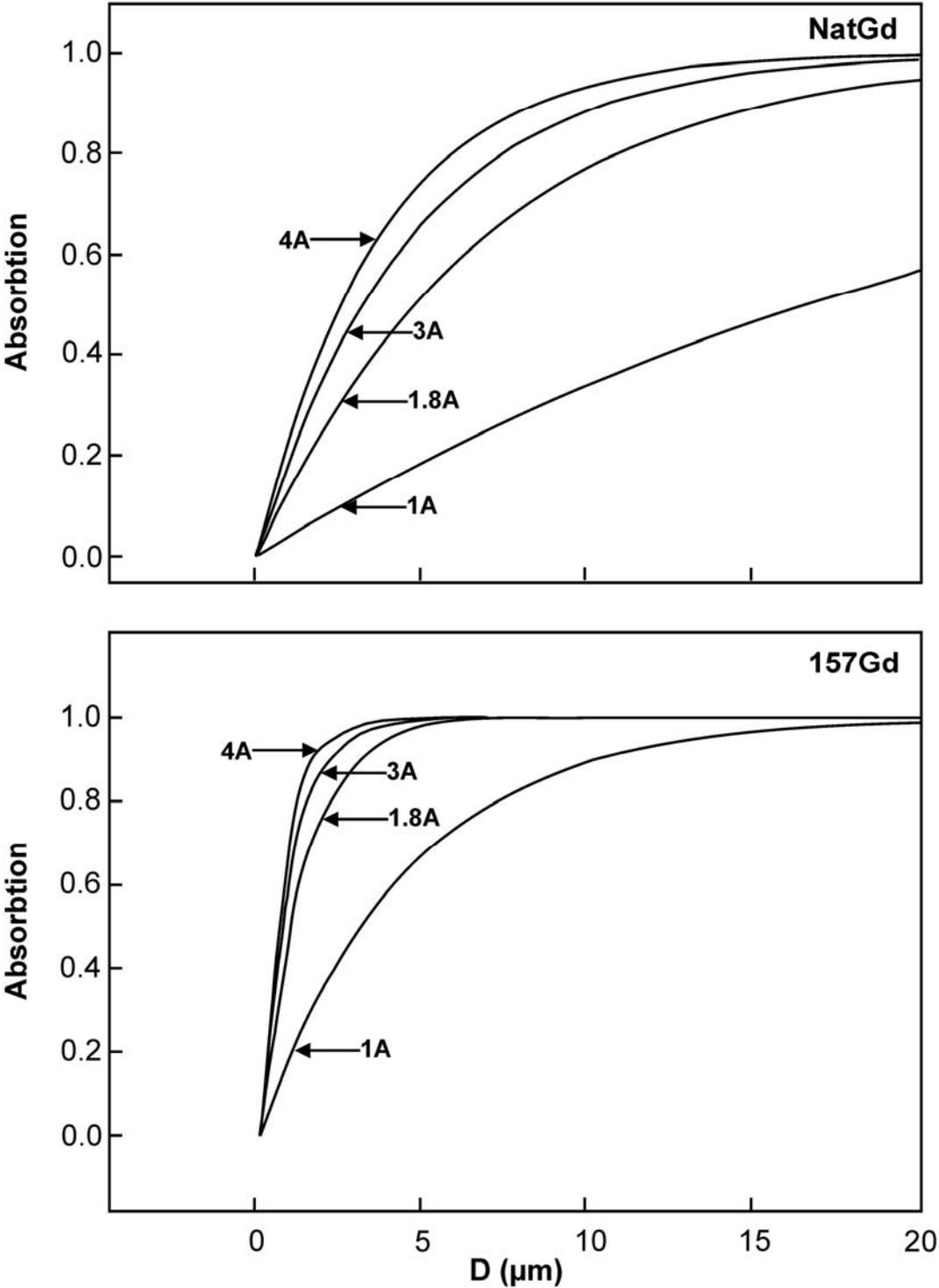

Fig. 5

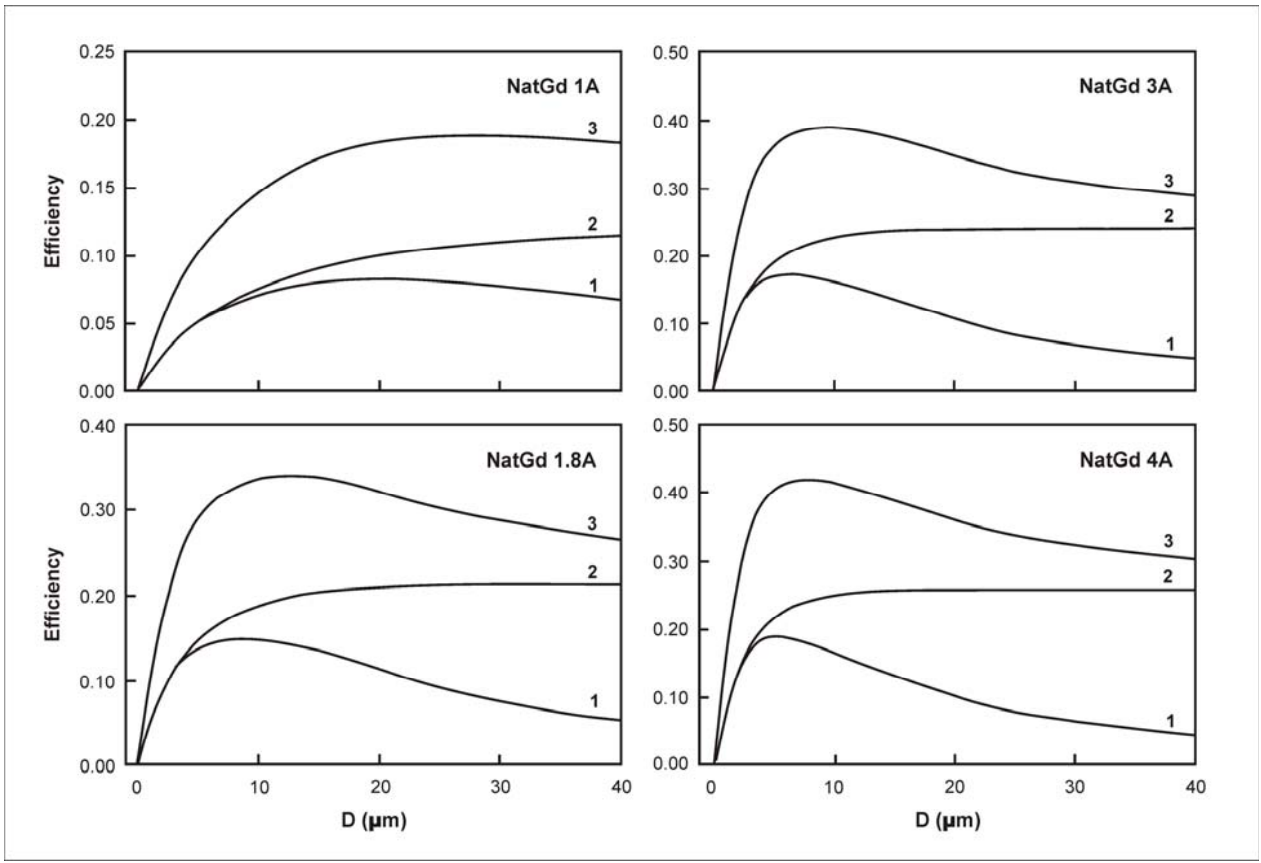

Fig. 6

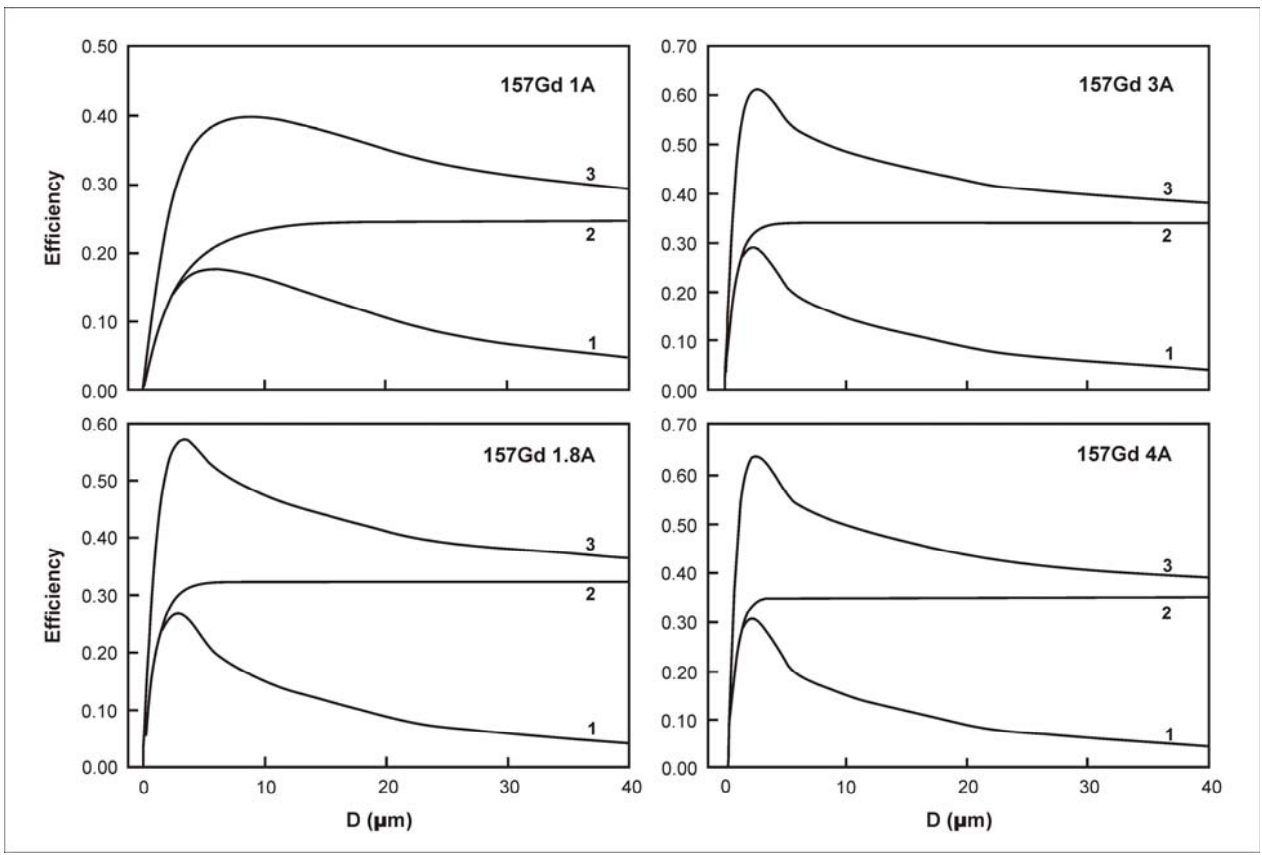

Fig. 7

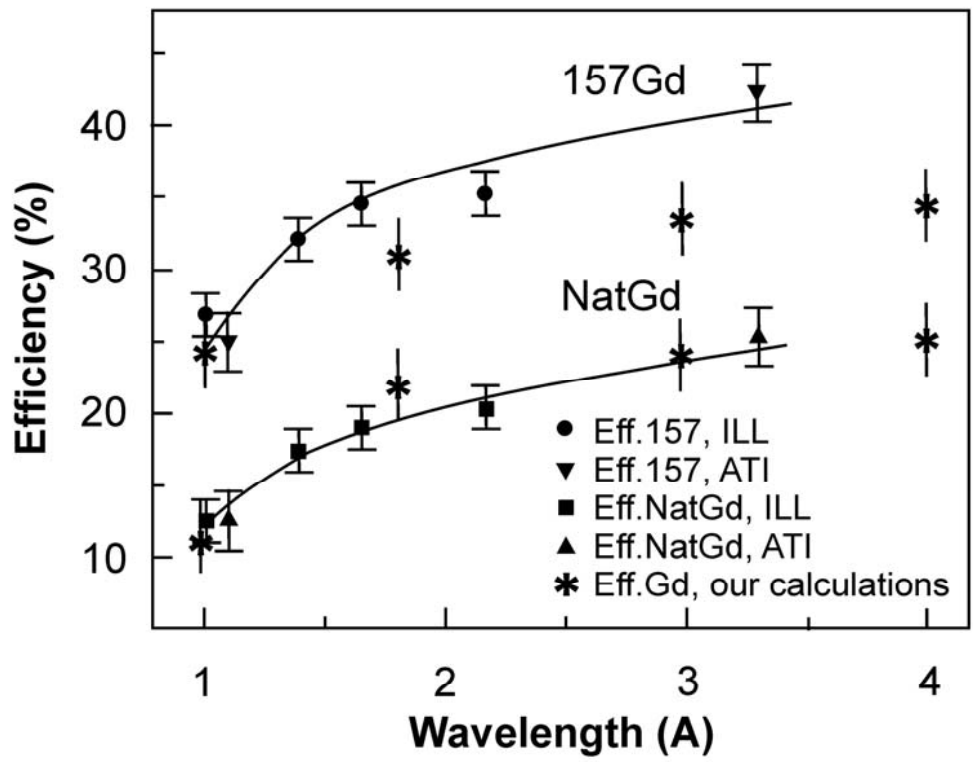

Fig.8